\documentclass[a4paper,fleqn,usenatbib]{mnras}

\usepackage{newtxtext,newtxmath}
\usepackage[T1]{fontenc}
\usepackage{ae,aecompl}
\usepackage{textcomp}
\usepackage[utf8]{inputenc}

\usepackage{mathtools,textcomp}
\usepackage{graphicx}	% Including figure files
\usepackage{subfigure}
\usepackage{amsmath}	% Advanced maths commands
\usepackage{amssymb}	% Extra maths symbols
\usepackage{nccmath}

\title[Characterising a highly obscured galaxy in ALPINE]{The ALPINE$-$ALMA [CII] Survey: On the nature of an extremely obscured serendipitous galaxy}

\author[M. Romano et al.]{M. Romano\thanks{E-mail: michael.romano@studenti.unipd.it}$^{1,2}$,
P. Cassata$^{1,2}$,
L. Morselli$^{1,2}$,
B. C. Lemaux$^{3}$,
M. B\'ethermin$^{4}$,\newauthor
P. Capak$^{5,6,7}$,
A. Faisst$^{5}$,
O. Le Fèvre$^{4}$,
D. Schaerer$^{8,9}$,
J. Silverman$^{10,11}$,
L. Yan$^{12}$,\newauthor
S. Bardelli$^{13}$,
M. Boquien$^{14}$,
A. Cimatti$^{15,16}$,
M. Dessauges-Zavadsky$^{8}$,
A. Enia$^{1,2}$,\newauthor
Y. Fudamoto$^{8}$,
S. Fujimoto$^{6,17}$,
M. Ginolfi$^{8}$,
C. Gruppioni$^{13}$,
N. P. Hathi$^{18}$,
E. Ibar$^{19}$,\newauthor
G. C. Jones$^{20,21}$,
A. M. Koekemoer$^{18}$,
F. Loiacono$^{13,15}$,
C. Mancini$^{1}$,
D. A. Riechers$^{22,23}$,\newauthor
G. Rodighiero$^{1,2}$,
L. Rodr\'iguez-Mu\~noz$^{1}$,
M. Talia$^{13,15}$,
L. Vallini$^{24}$,
D. Vergani$^{13}$,\newauthor
G. Zamorani$^{13}$,
and E. Zucca$^{13}$\\
\\
\centerline{\emph{\normalsize(Affiliations are listed at the end of the paper)}
}
}

\date{Accepted XXX. Received YYY; in original form ZZZ}

\pubyear{2020}

\begin{document}
\label{firstpage}
\pagerange{\pageref{firstpage}--\pageref{lastpage}}

\maketitle
%\clearpage
% Abstract of the paper
\begin{abstract}
We report the serendipitous discovery of a bright galaxy (Gal-A) observed as part of the ALMA Large Program to INvestigate [CII] at Early times (ALPINE). While this galaxy is detected both in line and continuum emission in ALMA Band 7, it is completely dark in UV/optical filters and only presents a marginal detection in the UltraVISTA $K_s$ band. We discuss the nature of the observed ALMA line, i.e. whether the emission comes from [CII] at $z\sim4.6$, or from high-J CO transitions at $z\sim2.2$. In the first case we find a [CII]-to-FIR luminosity ratio of $\mathrm{log(L_{[CII]}/L_{FIR})} \sim -2.5$, consistent with the average value for local star-forming galaxies (SFGs); in the second case, instead, the source would lie outside of the empirical relations between L$_\mathrm{CO}$ and $\mathrm{L_{FIR}}$ found in the literature. At both redshifts, we derive the star-formation rate (SFR) from the ALMA continuum, and the stellar mass ($\mathrm{M_{*})}$ by using stellar population synthesis models as input for LePHARE spectral energy distribution (SED) fitting. Exploiting our results, we believe that Gal-A is a \lq\lq Main-Sequence\rq\rq{} (MS), dusty SFG at $z=4.6$ (i.e. [CII] emitter) with $\mathrm{log(SFR/[M_{\odot} yr^{-1}])}\sim1.4$ and $\mathrm{log(M_{*}/M_{\odot})\sim9.7}$. This work underlines the crucial role of the ALPINE survey in making a census of this class of objects, in order to unveil their contribution to the global star-formation rate density (SFRD) of the Universe at the end of the Reionisation epoch.
\end{abstract}

% Select between one and six entries from the list of approved keywords.
% Don't make up new ones.

%%%%%%%%%%%%%%%%%%%%%%%%%%%%%%%%%%%%%%%%%%%%%%%%%%

%%%%%%%%%%%%%%%%% BODY OF PAPER %%%%%%%%%%%%%%%%%%

\section{Introduction}\label{sec:intro}

The last decades have seen dramatic advances in our knowledge of galaxy formation and evolution \citep{Giavalisco02,Renzini06,Silk12,Carilli13,Madau14,Naab17}. The global star-formation rate density (SFRD) has been found to rise during cosmic reionisation from $z \sim 10$, peak at $1 < z < 3$, and finally decrease by a factor of $\sim 10$ to the local Universe \citep{Lilly96,Bouwens11,Madau14}. Several studies suggest that, at all epochs, the bulk of the star-formation activity takes place in galaxies lying on the \lq\lq Main-Sequence\rq\rq{} (MS): a tight correlation between the star-formation rate (SFR) and the stellar mass ($\mathrm{M_{*}}$; \citealt{Daddi07,Rodighiero11,Speagle14,Santini17}). Therefore, we have indications that most of the stars in the Universe formed along the MS, at the peak of the SFRD at $z\sim2$. However, we are still trying to understand which are the main mechanisms responsible for the rapid increase of the SFRD at $z < 6$. One possible explanation is an increase in the gas fraction along with a rising in star-formation efficiency per unit mass, possibly driven by galaxy mergers \citep{Genzel15,Silverman15,Scoville16}.

At $z>3$ the cosmic SFRD is almost exclusively constrained by UV-selected samples \citep{Bouwens12a,Bouwens12b,Schenker13,Oesch15}, lacking information about the star formation obscured by dust. Rest-frame UV-selected galaxies must be corrected for dust absorption: wrong dust corrections can lead to large uncertainties on the SFR estimates and, consequently, to an incorrect picture of the star-formation history (SFH) of the Universe \citep{Gallerani10,Castellano14,Scoville15,Alvarez16}. At the same time, heavily dust-obscured star-forming galaxies (SFGs) may be completely missed by surveys probing rest-frame UV/optical emission. 

With the advent of new facilities, such as the Atacama Large Millimiter Array (ALMA), a population of faint, dusty SFGs has been confirmed at high redshift, e.g. sub-millimiter galaxies \citep{Dunlop04,Daddi09,Riechers10,Huang14,Santini16}, ALMA-only sources (e.g. \citealt{Williams19}), the extremely red objects selected with $H$ and IRAC colors (HIEROs galaxies) from \cite{Wang16}. While the bulk of these objects peaks at $2<z<3$, a significant tail of higher redshift, dusty galaxies without optical/near-infrared (NIR) detections is in place at $z>4$ \citep{Capak08,Daddi09,Riechers10,Walter12,Riechers13,Pavesi18}. For instance, \cite{Walter12} combined measurements from the IRAM Plateau de Bure Interferometer (PdBI) and the Jansky Very Large Array (VLA) to put constraints on the dust-obscured starburst HDF850.1, one of the first detected optical/NIR invisible galaxies. This source is at $z=5.18$ among an overdensity of galaxies at the same redshift, with a [CII]/FIR luminosity ratio comparable to that observed in local SFGs. In addition, most of these objects are often extreme starbursts, such as HFLS3. This source is confirmed to be at $z=6.34$ exploiting information from different molecular and atomic fine structure cooling lines and shows a large FIR luminosity (i.e. $\mathrm{L_{FIR}\sim2 \times 10^{13}}$ $\mathrm{L_\odot}$) and SFR $>10^3$ $\mathrm{M_{\odot}/yr}$ \citep{Riechers13}. 

An in-depth study of this elusive population of galaxies is necessary in order to complete the census of SFGs at high redshift contributing to the cosmic SFH as well as to better understand the early phases of the galaxy formation \citep{Blain02,Casey14}.

In this context, the ALMA Large Program to INvestigate [CII] at Early times (ALPINE; B\'ethermin et al. in prep.; \citealt{Faisst19}; \citealt{LeFevre19}) is going to improve our knowledge of the obscured star formation at $z>4$. It takes advantage of observations of the singly ionised carbon [CII] at 158\thinspace$\mathrm{\mu m}$ and its adjacent FIR continuum for a sample of 118 SFGs in the Cosmic Evolution Survey (COSMOS; \citealt{Scoville07a,Scoville07b}) and the Extended Chandra Deep Field South (E-CDFS; \citealt{Giavalisco04,Cardamone10}) fields. These sources are spectroscopically confirmed to be at $4 < z < 6$ with the Visible Multi-Object Spectrograph (VIMOS) at the Very Large Telescope (VLT; \citealt{LeFevre03,LeFevre15}) and with the DEep Imaging Multi-Object Spectrograph (DEIMOS) at the Keck II telescope \citep{Faber03,Hasinger18}. 

The [CII] line is one of the strongest lines in the FIR band (e.g. \citealt{Stacey91}) as it is one of the main coolants of the interstellar medium (ISM; \citealt{Carilli13}). Since it has a lower ionisation potential than neutral hydrogen (HI), i.e. 11.3 eV compared to 13.6 eV, this line can trace different gas phases, such as dense photodissociation regions (PDRs; \citealt{Hollenbach99}), neutral diffuse gas (e.g. \citealt{Wolfire03,Vallini15}), and diffuse ionised gas (e.g. \citealt{Cormier12}). In principle, in order to remove the ambiguity on the interpretation of the [CII] emission, the relative contribution of the various gas phases should be assessed. However, different studies suggest that the bulk of the [CII] emission arises from the external layers of molecular clouds heated by UV photons in PDRs \citep{Stacey91,Madden97,Kaufman99,Cormier15,Pavesi16}; thus, this line can be used as a tracer of star formation (e.g. \citealt{DeLooze14}; but see also \citealt{Zanella18} who suggest that [CII] is a better tracer of the molecular gas). Therefore, the combination of FIR continuum and UV measurements, together with the [CII] observations, will provide, at the redshift explored by the ALPINE survey, an estimate of the total (obscured and unobscured) star formation in these galaxies, corresponding to 80-95$\%$ of the cosmic SFRD at $4<z<6$ \citep{Casey12,Bouwens16,Capak15,Aravena16,Novak17}. 
The remaining 5-20$\%$ of star formation which is not traced by UV data is yielded by a free blind survey covering an additional area of 25 arcmin$^2$ beyond the targeted sources, where many galaxies have been serendipitously detected so far (Loiacono et al. in prep.). Among these, several sources are invisible in the optical bands. The study of these objects is crucial for obtaining a robust estimate of the total SFRD at $z > 4$ and for characterising the overall population of the high-redshift SFGs.

In this work, we discuss the nature of a galaxy (hereafter, Gal-A) randomly discovered in the field of the ALPINE target DEIMOS$\_$COSMOS$\_$665626 (hereafter, DC$\_$665626). The galaxy has a spatial offset of $\sim 6$ arcsec (1 arcsec is $\sim 7$ kpc at $z = 4.583$, the redshift of the target) from DC$\_$665626; it does not show any optical counterpart at the position of the emission detected with ALMA and, for this reason, its nature results to be ambiguous. Besides, since Gal-A is the brightest galaxy detected in line emission among all those having no optical counterpart and serendipitously observed in ALPINE so far (Loiacono et al. in prep.), this work can be exploited as a benchmark for future analysis on these types of sources.

The paper is organised as follows: in Section 2 we introduce the available data we have for Gal-A, and explain the methods used to analyse this source. We present the results in Section 3 and discuss them in Section 4, trying to constrain the nature of the galaxy. Summary and conclusions are provided in Section 5. 

Throughout this paper, we assume a $\Lambda$CDM cosmology with $\mathrm{H_0}$ = 70 km/s/Mpc, $\mathrm{\Omega_m}$ = 0.3, and $\mathrm{\Omega_\Lambda}$ = 0.7 \citep{Planck18}. We furthermore use a \cite{Chabrier03} initial mass function (IMF) and AB magnitudes. 

%%%%%%%%%%%%%%%%%%%%%%%%%%%%%%%%%%%%
\section{Observations and Data Reduction}

\subsection{ALMA data}
DC$\_$665626 has been observed with ALMA in Band 7 ($\nu_{\mathrm{obs}}=[275-373]$ GHz) on 25 May 2018 (Cycle 5; Project 2017.1.00428.L, PI O. Le F\`evre) using 45 antennas with the C43-2 array configuration (with a maximum baseline of $\sim$ 250 m). The on-source integration time is 16 minutes, with a total elapsed time of 37 minutes.

The spectral setup consists of two sidebands with a frequency range of $\mathrm{\Delta_{\nu}^{l} \simeq [339-343]}$ GHz and $\mathrm{\Delta_{\nu}^{u} \simeq [351-355]}$ GHz for the lower and upper sidebands, respectively. Each sideband is made up of two spectral windows (SPWs) of width 1.875 GHz, each of which containing 128 channels 15.625 MHz wide (the sidebands overlap for 7 channels), with a typical rms of 0.6 mJy beam$^{-1}$ per channel. The flux and phase are calibrated using the standard calibrators J1058+0133 and  J0948+0022, respectively.

The data are analysed using standard pipelines for ALMA reduction included in the software CASA \citep{McMullin07}, version 5.4.0. The imaging is obtained running the \texttt{TCLEAN} task on the visibilities, setting a threshold of 3$\mathrm{\sigma_{rms}}$ on the noise level when cleaning the data (where $\mathrm{\sigma_{rms}}$ is obtained from the dirty image), and with a natural weighting scheme to increase the sensitivity.

\subsection{Identification of the serendipitous source}
As part of the COSMOS field \citep{Scoville07a,Scoville07b}, which is one of the most thoroughly studied regions of the sky so far, multi-wavelength data are available for the whole ALPINE sample, including high-resolution Hubble Space Telescope (HST) imaging \citep{Koekemoer07,Koekemoer11}, and photometry from the Canada-France-Hawaii Telescope (CFHT), the Spitzer telescope and other facilities \citep{Capak07,Laigle16}. Spectroscopic redshifts are available from large optical spectroscopic campaigns at the VLT (VUDS; \citealt{LeFevre15}) and Keck (DEIMOS; \citealt{Hasinger18}). Multi-band photometry and spectroscopic data allow to build spectral energy distributions (SEDs) and to derive robust parameters including SFRs and stellar masses through SED-fitting \citep{Faisst19}. Through this analysis, we find that DC$\_$665626 has log$\mathrm{(M_{*}/M_{\odot})} = 9.21^{+0.16}_{-0.18}$, log$\mathrm{(SFR/[M_{\odot} yr^{-1}]) = 0.71^{+0.29}_{-0.18}}$, and a spectroscopic redshift of $z_\mathrm{spec} = 4.583$, obtained from Ly$\alpha$ emission and ISM absorption lines in the observed-frame optical spectrum.

\begin{figure}
	\includegraphics[width=\columnwidth]{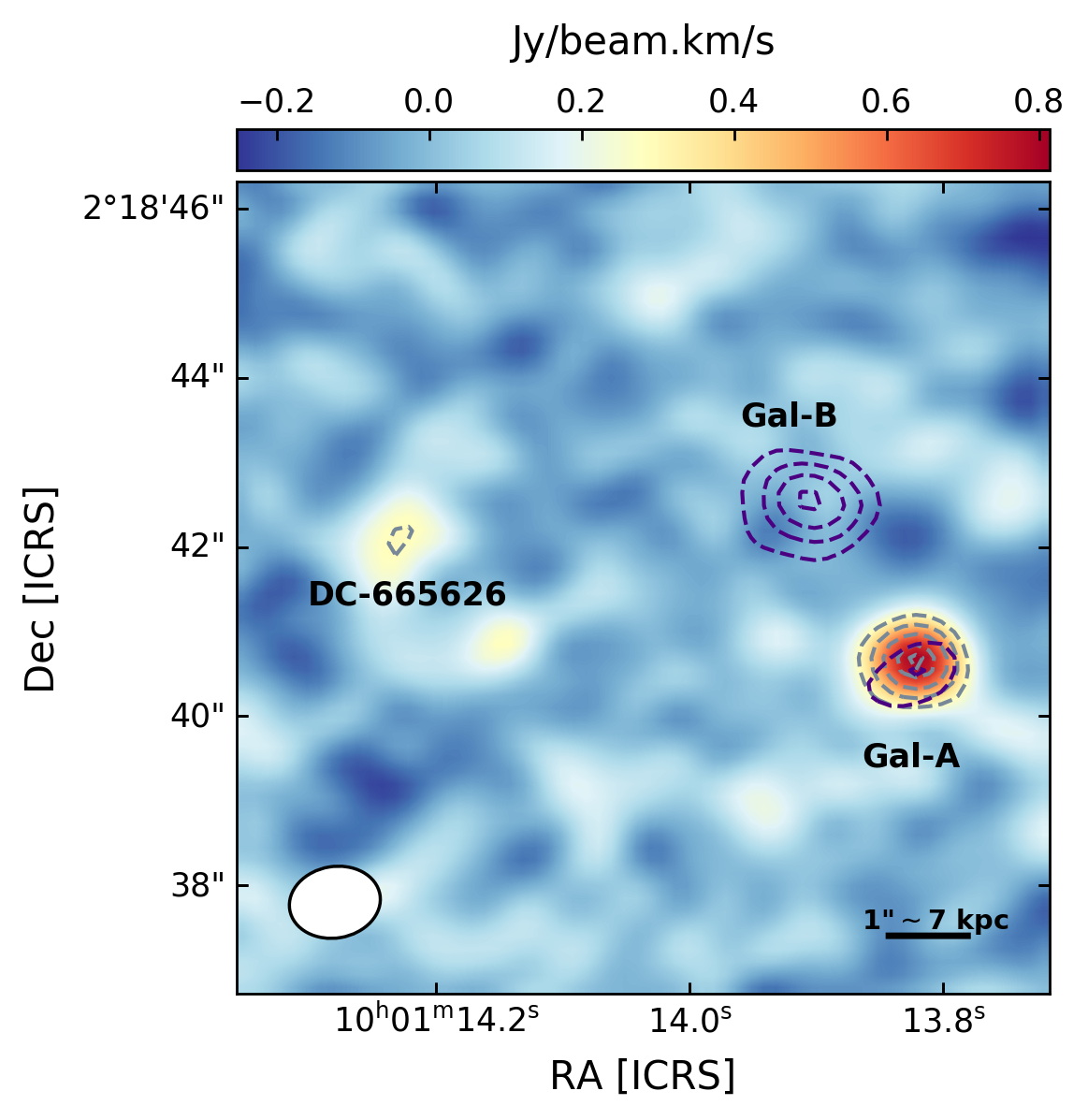}
    \caption{Continuum-subtracted moment-0 map of Gal-A. The ALPINE target DC$\_$665626, Gal-A and Gal-B are labelled. Line and continuum emissions are shown with grey and purple contours starting from $4\sigma$ and $3\sigma$ (at step of 2$\sigma$), respectively. The white ellipse in the bottom left corner is the synthesized beam.}
    \label{fig:maps}
\end{figure}

\begin{figure}
	\includegraphics[width=\columnwidth]{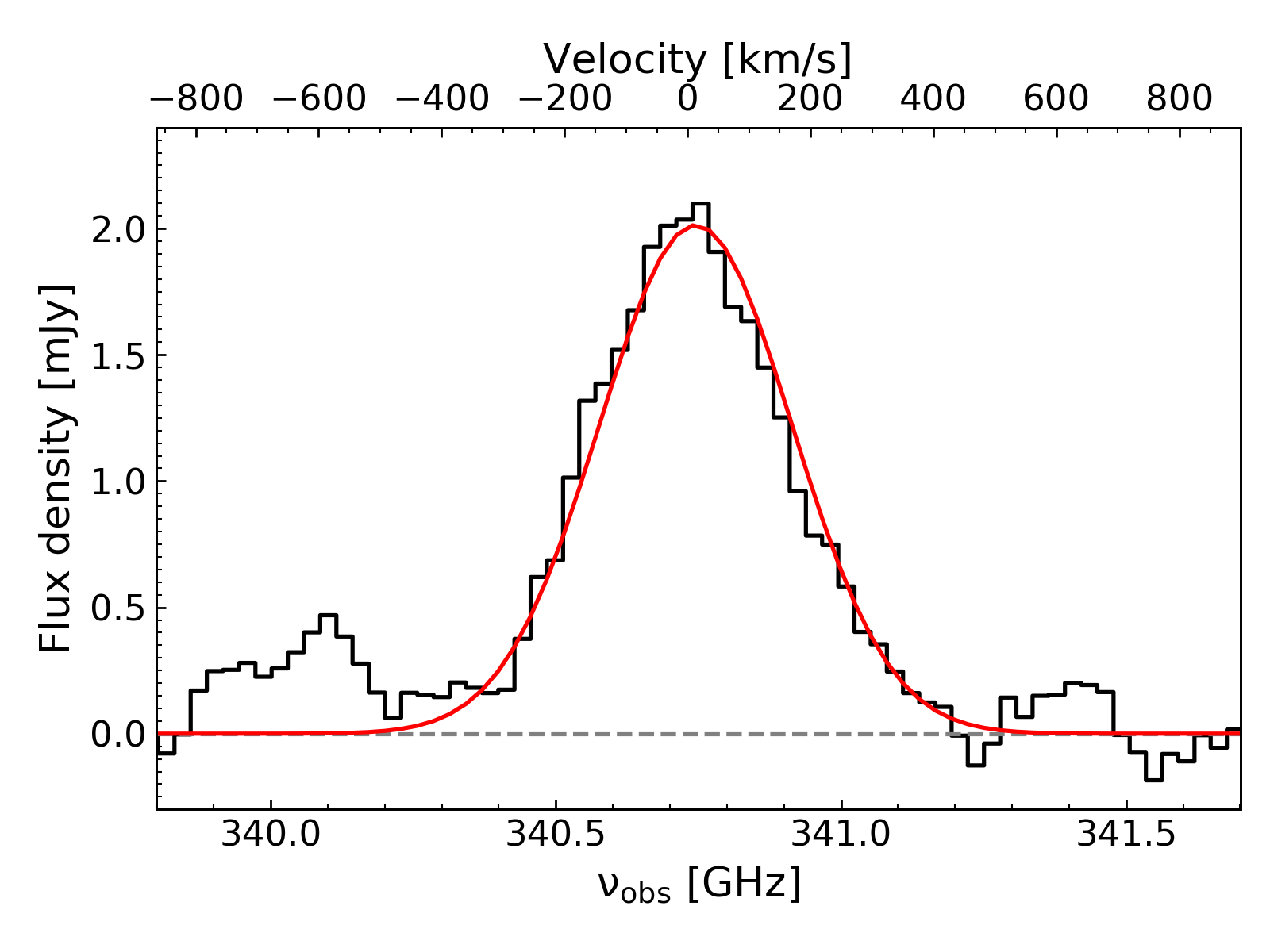}
    \caption{Emission line flux at the position of Gal-A (black histogram) as a function of the observed frequency. The solid red curve represents the gaussian fit on the line. The dashed grey line marks the zero-flux level. Also shown is the velocity offset on the top axis.}
    \label{fig:CII_spec}
\end{figure}

\begin{figure*}
     \begin{center}
     \includegraphics[width=1.7\columnwidth]{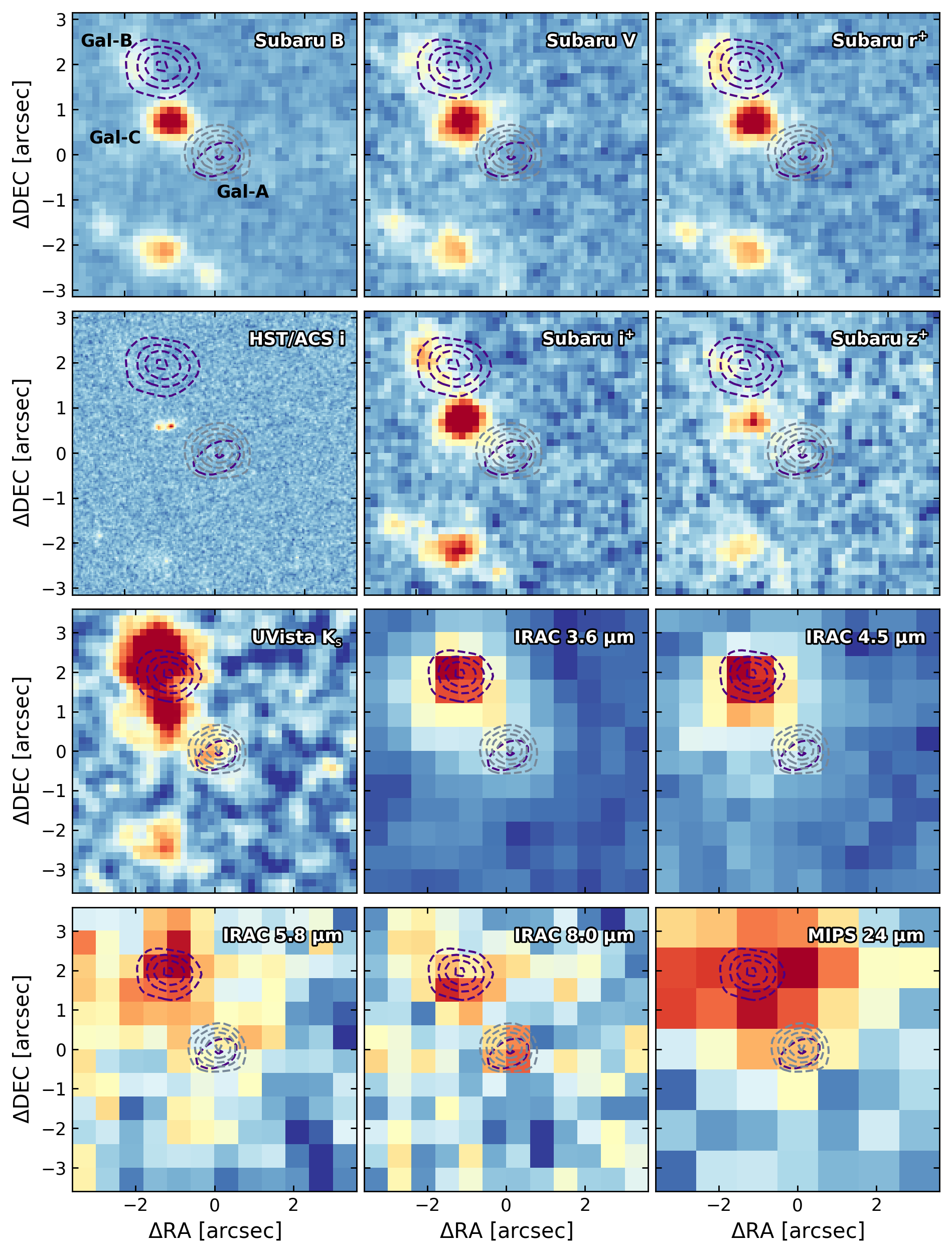}
    \end{center}
    \caption{Cutouts centred on Gal-A in different photometric filters, from HST/ACS \citep{Koekemoer07} and Subaru, UltraVISTA and Spitzer \citep{Capak07,Laigle16}. Grey and purple contours are $>3\sigma$ line and continuum emissions (at step of 2$\sigma$), respectively. Gal-A, Gal-B, and Gal-C are labelled in the upper-left plot of the figure. Wavelengths increase from the upper-left to the bottom-right corner.}
   \label{fig:cutouts}
\end{figure*}

Since the ALPINE target DC$\_$665626 is at $z_\mathrm{spec}=4.583$, the [CII] emission from this source ($\mathrm{\nu_{rest} = 1900.54}$ GHz) is expected to be redshifted at around $\mathrm{\nu_{obs}=340.42}$ GHz, falling inside the lower sideband of the observed ALMA spectrum. When we inspect the cube, together with the [CII] emission coming from DC$\_$665626 (at 4.4$\sigma$; B\'ethermin et al. in prep.), we identify a more significant emission with a spatial offset of $\sim 6$'' ($\sim 40$ proper kpc at $z\sim4.6$) with respect to the ALPINE target. We refer to the source of this emission as Gal-A (RA: 10:01:13.82, Dec: +02:18:40.66), that is detected both in continuum and in line emission at 5$\sigma$ and 12$\sigma$, respectively. Fig. \ref{fig:maps} shows the continuum-subtracted moment-0 map of Gal-A (see section \ref{sec:analysis}). Also shown are the synthesized beam with a size of 1.08'' $\times$ 0.85'' at P.A. = - 80$^{\circ}$, and another galaxy (Gal-B) detected at 9$\sigma$ in continuum only northwards of the offset emission ($\sim$ 2'' away from Gal-A when considering the peak position of the two emissions).

We show in Fig. \ref{fig:CII_spec} the spectrum of the emission line observed at the position of Gal-A; it is extracted from a circular region 2'' wide, including the 2$\sigma$ contours from the moment-0 map of the source. Using the \textit{spectral profile tool} within the CASA viewer, we fit the line profile with a Gaussian function finding a full width at half maximum (FWHM$_\mathrm{line}$) of $308 \pm 34$ km/s and a peak frequency at $\mathrm{\nu_{peak}} = 340.76$ GHz.

Though DC$\_$665626 is detected in [CII] in spatial coincidence with its UV emission, we consider the possibility that the emission centred at the position of Gal-A is connected with that of the ALPINE target. The displacement between [CII] and UV/Ly$\alpha$ emission has already been observed in high-redshift galaxies (\citealt{Gallerani12,Willott15}; Cassata et al. in prep.; but see also \citealt{Bradac17}). It is also reproduced by radiative transfer simulations as a consequence of the strong stellar feedback which could quench the [CII] emission in the central region of the galaxies, allowing it to arise mostly from infalling or satellite clumps of neutral gas around them \citep{Vallini13,Maiolino15}. However, these models predict spatial offsets up to $\sim 1-2$ arcsec ($\sim 7-14$ kpc at the redshift of the target), well below the offset that we measure in this case ($\gtrsim 6$ arcsec). Therefore, we exclude that the observed ALMA emission at the position of Gal-A is directly linked to DC$\_$665626.

\subsection{Multi-wavelength photometry of Gal-A}
As Gal-A lies in the COSMOS field \citep{Laigle16}, we exploit all the available multi-wavelength photometry in order to identify the counterpart associated with the discovered emission. In Fig. \ref{fig:cutouts} we present some cutouts centred on this galaxy in different photometric filters, from UV to FIR. Gal-B, that in the COSMOS2015 catalogue \citep{Laigle16} has a photometric redshift $z = 2.249^{+0.223}_{-0.151}$, is visible in most of the photometric bands. Another foreground galaxy, labelled Gal-C in Fig. \ref{fig:cutouts}, is well detected in the images from optical to NIR wavelengths, and has $z = 2.021^{+0.123}_{-0.116}$ in COSMOS2015. Conversely, Gal-A is not clearly visible in any optical filter except for the UltraVISTA $K_s$ band, even if it is not listed as a detection in the UltraVISTA DR4 catalogue \citep{McCracken12}. 

More in detail, to reproduce the SED of Gal-A, we use observations in $u^{*}$ band from MegaCam on CFHT, as well as the $B$, $V$, $r^{+}$, $i^{+}$, and $z^{++}$ filters from Suprime-Cam on Subaru, in order to set an upper limit to the optical emission of the source. NIR constraints come from the $J$, $H$, and $K_s$ bands from VIRCAM on the VISTA telescope. Finally, we obtain information on the SED up to $\sim8$ $\mu$m in the observed-frame from the IRAC channels on Spitzer. For each band, we centre a fixed aperture of 1.4'' of diameter on Gal-A (enclosing the 3$\sigma$ contours of the emission line detected by ALMA) and estimate its flux. The rms is computed as the average rms within several apertures (of 1.4'' of diameter) placed in different regions of the sky, close to the source but away from evident emission. As expected, we don't find any significant detection of our source in the optical bands. Some marginal detections are present in the $B$, $r^{+}$, and $i^{+}$ filters, but they are all below $2\sigma$ and could be partially contaminated by the emission of Gal-C. For this reason, we consider the fluxes measured in these bands as upper limits. The same argument applies to the VISTA filters, except for the $K_s$ band in which, as mentioned above, a faint emission arises at the position of Gal-A. Making use of \texttt{SExtractor} \citep{Bertin96}, we manage to deblend the analysed galaxy from the other two nearby sources obtaining an estimate of its apparent magnitude in this band. Through this analysis, Gal-A is detected at $\sim$ 2.3$\sigma$, with an AB magnitude $K_s = (24.8 \pm 0.5$), which is very close to the corresponding 3$\sigma$ limiting magnitude of $\sim25$ from the UltraVISTA DR4 catalogue.

Finally, a weak emission seems to arise at the position of Gal-A in the IRAC bands. However, as shown in Fig. \ref{fig:cutouts}, this could be partially contaminated by the emission of the two nearby galaxies at $z\sim2$ in the 3.6 and 4.5 $\mu$m bands, while it seems to emerge from the background at 8.0 $\mu$m, where Gal-B and Gal-C become fainter. We find that, in the \textit{Rainbow} catalogue \citep{Perez08,Barro11a,Barro11b}, Gal-B and Gal-C have been deblended in all the four IRAC channels using the Subaru $r$ band as a prior for the two sources, while no counterpart of Gal-A is present.   

In order to extract the photometric information on Gal-A from the IRAC bands, we attempt a deblending procedure using the 2D GALFIT fitting algorithm \citep{Peng02}. We model Gal-B and Gal-C as point-like sources, using their optical positions and deblended fluxes from \textit{Rainbow} as a first guess, and considering for each IRAC channel its typical PSF ($\sim 2''$). To obtain the Gal-A flux in each channel, we perform aperture photometry at the position of Gal-A in the residual maps\footnote{As we use a fixed aperture of 1.4'' (which is smaller than the typical PSF of the IRAC channels) centred on Gal-A to obtain the photometry in each band, we apply aperture corrections to estimate the total fluxes in the IRAC filters. In particular, we divide the flux measured in these bands by 0.61, 0.59, 0.49 and 0.45, going from 3.6 to 8.0 $\mu$m}. We are aware that with this procedure we may underestimate the flux of Gal-A in the IRAC channels as we are spreading the global flux of the three components on only two sources. To account for this, when performing SED-fitting (see Section \ref{section:mstar}) we decide to consider IRAC fluxes ranging between the deblended (lower) and blended (higher) values. We find, however, that our conclusions do not depend on this assumption; in fact, we obtain similar results when using the deblended fluxes in the SED-fitting. As an alternative approach, we tried to fit a three-components model leaving as a free parameter the flux corresponding to Gal-A and using the ALMA continuum peak position as a prior. However, probably due to the small distance between the galaxies, the code is not able to perform the fit. Table \ref{tab:phot} summarises the photometric information we obtain for Gal-A; this is exploited in section \ref{section:mstar} to estimate the stellar mass of this galaxy. 

\begin{table}
\begin{center}
\begin{tabular}{c c c c c}
\hline
\hline
Instrument & Filter & Central $\mathrm{\lambda}$ & Observed flux\\
/Telescope & & [$\mu$m] & [$\mu$Jy]\\
\hline
MegaCam/CFHT & $u^{*}$ & 0.3783 & $<1.93\times10^{-2}$ \\ \hline
Suprime-Cam & $B$ & 0.4458 & $<4.12\times10^{-2}$\\
/Subaru & $V$ & 0.5478 & $<6.90\times10^{-2}$\\ 
& $r^{+}$ & 0.6289 & $<6.88\times10^{-2}$\\
& $i^{+}$ & 0.7684 & $<9.32\times10^{-2}$\\
& $z^{++}$ & 0.9037 & $<2.92\times10^{-1}$\\
\hline
VIRCAM & $J$ & 1.2495 & $<3.70\times10^{-1}$\\
/VISTA & $H$ & 1.6553 & $<5.22\times10^{-1}$\\
& $K_s$ & 2.1640 & (4.25$\pm$1.85)$\times10^{-1}$\\
\hline
IRAC/Spitzer & ch1 & 3.5634 & $<1.10$\\
& ch2 & 4.5110 & $<1.36$\\
& ch3 & 5.7593 & $<2.12$\\
& ch4 & 7.9595 & $<4.27$\\
\hline
\hline
\end{tabular}
\caption{Summary of available data for Gal-A in each photometric band used for the SED-fitting (see section \ref{section:mstar}). The first two columns are the instruments (with relative telescopes) and filters used. Central wavelength is the mean wavelength weighted by the transmission of the filter. In the last column, the fluxes (wich are all 2$\sigma$ upper limits except for the $K_s$ band) of Gal-A are shown. For the IRAC channels, we report the upper limits obtained by measuring the flux of Gal-A before the deblending procedure (see text). Except for the $K_s$ detection (which is obtained with \texttt{SExtractor}), all the estimated photometry is directly obtained from the maps with an aperture of 1.4'' of diameter centered on Gal-A.}
\label{tab:phot}
\end{center}
\end{table}

\begin{table*}
\begin{center}
\begin{tabular}{c c c c c c}
\hline
\hline
& $\mathrm{\nu_{rest}}$ &  $z_\mathrm{gal}$ & $\mathrm{log(L_{line})}$ & $\mathrm{log(L_{FIR})}$ & log(SFR) \\
& [GHz] & & [L$_{\odot}$] & [L$_{\odot}$] & [M$_{\odot}$/yr] \\
\hline
CO(9-8) & 1036.9 & 2.043 & 8.04 $\pm$ 0.04 & 11.44 $\pm$ 0.50 & 1.45 $\pm$ 0.50 \\
CO(10-9) & 1152.0 & 2.381 & 8.20 $\pm$ 0.04 & 11.42 $\pm$ 0.50 & 1.43 $\pm$ 0.50 \\
{[CII]} & 1900.5 & 4.577 & 8.88 $\pm$ 0.04 & 11.38 $\pm$ 0.50 & 1.38 $\pm$ 0.50 \\
\hline
\hline
\end{tabular}
\caption{Summary of the physical parameters estimated for the three possible emission lines attributed to Gal-A. The first three columns report the considered emission line, its rest-frequency emission, and the redshift $z_\mathrm{gal}$ derived using the observed peak frequency, respectively. The fourth and fifth columns list the line luminosity ($\mathrm{L_{line}}$) and the total infrared luminosity ($\mathrm{L_{FIR}})$ for each emission lines, respectively. Finally, the last column report the SFRs, directly computed from the FIR luminosities following \citealt{Kennicutt98}.}
\label{tab:param}
\end{center}
\end{table*}

\subsection{Analysis of the serendipitous source}
\label{sec:analysis}
Since Gal-A shows no optical counterpart, we do not know a priori the nature of the emission line; it could be [CII] emission at a similar redshift of DC$\_$665626 (i.e. $z\sim4.6$), but also high-J CO transitions are expected ($\mathrm{J_{up}} > 3$) at the observed frequencies in ALMA Band 7, although at lower redshift \citep{Carilli13}.

In this work, we consider only the two high-J CO transitions with $\mathrm{J_{up}}=9,10$ which fall into the SPW of observation at $z\gtrsim2$. Indeed, \cite{Ilbert13} claim that galaxies at $z<2$ (corresponding in our case to lower CO transitions) should be more easily detected in UV/optical filters, with a fraction always greater than 95\% of sources detected in at least four photometric bands, from UV to NIR \citep{Ilbert06}. Therefore, if our source was at $z<2$, we would expect it to be visible in the optical bands shown in Fig. \ref{fig:cutouts}. 

For these reasons, in this work we discuss the nature of Gal-A considering three transitions as possible interpretations for the observed emission: [CII] at $\mathrm{\nu_{rest}}=1900.5$ GHz, CO(9-8) at  $\mathrm{\nu_{rest}}=1036.9$ GHz, and CO(10-9) at $\mathrm{\nu_{rest}}=1152.0$ GHz. As the observed emission line has a peak frequency of 340.76 GHz, Gal-A would be at redshift $z_\mathrm{gal} = 4.577$, $z_\mathrm{gal} = 2.043$ and $z_\mathrm{gal} = 2.381$ for [CII], CO(9-8) and CO(10-9), respectively. Table \ref{tab:param} lists the considered transitions and their rest frequencies, as well as the corresponding redshift for Gal-A in the three cases.  

To estimate the intensity of the line and continuum emissions from Gal-A, we separate these components using the CASA \texttt{IMCONTSUB} task; in particular, giving in input all the channels in the SPWs free of the emission line, this task creates a continuum map of the source and a continuum-subtracted cube. We then select all the consecutive channels having emission above 1$\mathrm{\sigma_{spec}}$ (i.e. the rms estimated from the line spectrum) encompassing the emission line in order to compute the moment-0 map with the CASA \texttt{IMMOMENTS} task. 

\begin{figure*}
\begin{center}
	\includegraphics[width=13cm,angle=0]{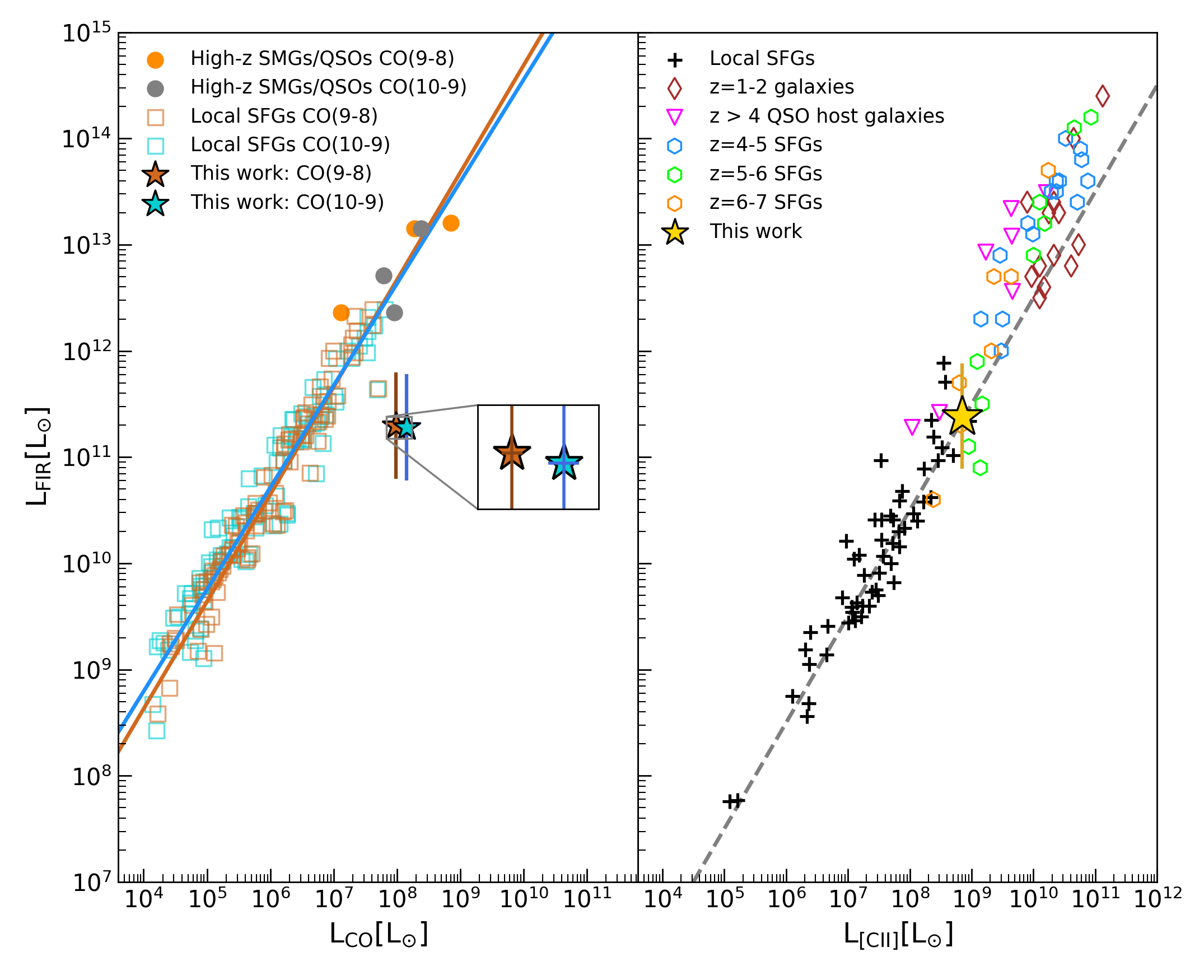}
    \caption{Left panel: empirical relations between CO(9-8) (solid brown line), CO(10-9) (solid blue line) and FIR luminosity \citep{Liu15} with overlaid the values for individual local galaxies as brown and blue open squares, respectively (Liu et al. 2015, private communication). The two stars are the values found for Gal-A in this work (same color legend). Error bars are estimated by propagating the error of the line flux on L$\mathrm{_{CO}}$, and assuming a variation of 0.5 dex for $\mathrm{L_{FIR}}$. Also shown are the values obtained for high-redshift sub-mm galaxies/quasi-stellar objects (QSOs) as the orange and grey filled circles, in case of CO(9-8) and CO(10-9) transitions, respectively \citep{Carilli13,ALMA15,Carniani19}. Right panel: [CII] as a function of FIR luminosity for several kinds of objects at different redshifts. Black crosses are local SFGs \citep{Malhotra01b}; brown diamonds are $z = 1-2$ galaxies, including starburst- and AGN-dominated sources \citep{Stacey10}; magenta triangles are $z = 4.1-7.1$ QSO host galaxies \citep{Pety04, Maiolino05, Iono06, Maiolino09, Wagg10, Willott13}; $z = 4-7$ SFGs are the cyan, green, and orange hexagons \citep{Lagache18}. The dashed grey line represents the average [CII]-to-FIR ratio for local galaxies \citep{Ota14}. The yellow star shows the position of our source. Error bars are estimated by propagating the error of the line flux on L$_\mathrm{[CII]}$, and assuming a variation of 0.5 dex for $\mathrm{L_{FIR}}$.}
    \label{fig:LCII}
\end{center}
\end{figure*}

The line and continuum fluxes are computed using the CASA \texttt{IMFIT} task. We define a region surrounding the emissions and then select only the pixels with a flux density larger than 2$\mathrm{\sigma}$: since the size of the emission region is comparable with the clean beam size, we assume that the source is unresolved and we take the peak flux as the total flux. We obtain $\mathrm{S_{cont}} = 245 \pm 24$ $\mu$Jy and $\mathrm{S_{line}\Delta v = 1.19 \pm 0.11}$ Jy km/s for the continuum and the line, respectively.

We derive the total infrared (between 8 and 1000 $\mu$m) luminosity of the source, in the three cases, assuming a shape of its SED from \cite{Magdis12}, and normalizing its flux to $\mathrm{S_{cont}}$, which is the observed flux at $\sim 845-880$ $\mu$m; according to \cite{Kennicutt98}, this luminosity also provides a good estimate of the obscured SFR. We obtain $\mathrm{log(L_{FIR}/L{_\odot})}=11.38\pm0.5$ in case of [CII] emission, $\mathrm{log(L_{FIR}/L{_\odot})}=11.44\pm0.5$ for CO(9-8), and $\mathrm{log(L_{FIR}/L{_\odot})}=11.42\pm0.5$ for CO(10-9) emissions. The uncertainties on the FIR luminosities are calculated by adding in quadrature the error on the continuum flux ($\sim0.04$ dex, which directly affects the $\mathrm{L_{FIR}}$ estimates), and a systematic error of 0.5 dex which takes into account possible variations in the luminosity caused by different FIR SED templates; as can be seen, this latter term dominates over the uncertainty on the flux. Following Eq. (4) in \cite{Kennicutt98}, these FIR luminosities translate into SFRs ranging from 24 to 28 $\mathrm{M_{\odot}/yr.}$\footnote{We scale the SFR from Salpeter to Chabrier IMF by dividing by 1.7 (e.g. \citealt{Zahid12}).} Finally, we estimate the line luminosities as in \cite{Solomon92} using the following relation:

\begin{ceqn}
\begin{equation}
\mathrm{L_{line} = 1.04 \times 10^{-3} \hspace{0.5mm} S_{line}\Delta v \hspace{0.5mm} D^2_L \hspace{0.5mm} \nu_{obs} \hspace{1mm} [L_{\odot}]},
\label{eq:L_CII}
\end{equation}
\end{ceqn}

\noindent where $\mathrm{D_L}$ is the luminosity distance of the source in Mpc, and $\mathrm{\nu_{obs}}$ the observed peak frequency in GHz. We thus obtain $\mathrm{log(L_{[CII]}/L{_\odot})}=8.88\pm0.04$, $\mathrm{log(L_{CO}/L{_\odot})}=8.04\pm0.04$ for CO(9-8) and $\mathrm{log(L_{CO}/L{_\odot})}=8.20\pm0.04$ for CO(10-9), where the uncertainties are computed by propagating the line flux error on the above equation. Table \ref{tab:param} reports all the above-mentioned physical quantities computed for Gal-A.

%%%%%%%%%%%%%%%%%%%%%%%%%%%%%%%%%%%%%

\section{Results}

\subsection{On the nature of the serendipitous source}
With the only information of the ALMA Band 7 line and continuum, and with no detections in optical bands, unveiling the nature of Gal-A is a challenging task. We use the physical quantities estimated in section \ref{sec:analysis} to deduce plausible conclusions on this source.

Fig. \ref{fig:LCII} (left panel) shows the correlation between L$\mathrm{_{CO}}$ (for the (9-8) and (10-9) transitions) and $\mathrm{L_{FIR}}$ for a compilation of SFGs in literature, together with the expected position of Gal-A; the respective best-fitting lines on the individual data are also shown (solid lines, \citealt{Liu15}). It is worth noting that the reported values are for local galaxies, spanning a FIR luminosity range between $\sim 10^8-10^{12}$ L$_{\odot}$. However, the empirical correlations continue to apply even including high-redshift galaxies (open diamonds in the figure); in this case indeed, as shown in \cite{Liu15}, the results of the fit do not significantly change. We then note that the computed $\mathrm{L_{FIR}}$ of \cite{Liu15} are integrated between $40-400$ $\mu$m, which is a smaller range with respect to the one adopted in this paper. In order to take this difference into account, we rescale the FIR luminosities of Gal-A in Fig. \ref{fig:LCII} to the same integration interval as in \cite{Liu15}, for consistency ($\mathrm{L_{FIR}^{8-1000}/L_{FIR}^{40-400} \sim 1.4}$, on average). It can be seen that, for both possible CO transitions, our galaxy would be an outlier of the empirical relations found by \cite{Liu15}, if it was at $z\sim2$. However, considering the large uncertainties on $\mathrm{L_{FIR}}$ (i.e. 0.5 dex), Gal-A could still be part of the lower envelope of local SFGs in the figure, tracing high-density regions ($\mathrm{n_{H2,crit}}\sim10^5-10^6$ $\mathrm{cm}^{-3}$; \citealt{Carilli13}) where star formation may occur.

In the right panel of Fig. \ref{fig:LCII} we plot the [CII] luminosity as a function of $\mathrm{L_{FIR}}$ in case Gal-A was a [CII] emitter at $z \sim 4.6$, along with other results from several authors for different types of objects (e.g. \citealt{Malhotra01b}, \citealt{Stacey10}). Our source perfectly sits on the local SFGs relation, with $\mathrm{log(L_{[CII]}/L_{FIR})} \sim -2.5$; possibly, this galaxy may belong to the high-redshift SFGs population which extends to $\mathrm{log(L_{FIR}/L_{\odot})} \sim 11$. As previously said, the [CII] line is mostly produced by UV radiation field in star-forming regions (e.g. \citealt{Cormier15}), then it can trace the SFR. Therefore, as the FIR emission marks out the SFR of a source, the relation between [CII] luminosity and $\mathrm{L_{FIR}}$ translates into a correlation between $\mathrm{L_{[CII]}}$ and the SFR of a galaxy; Gal-A follows this relation, not showing the typical [CII] deficit which arises at $\mathrm{L_{FIR}>10^{11}}$ $\mathrm{L_{\odot}}$ \citep{Luhman98,Malhotra01a,Luhman03,Lagache18}. 

These results suggest that our source, randomly detected in the DC$\_$665626 field, may more likely be a strongly obscured [CII] emitter at high redshift. However, to validate this hypothesis, more data are needed.

\subsection{Estimate of the stellar mass}
\label{section:mstar}

\begin{figure}
\begin{center}
	\includegraphics[width=\columnwidth]{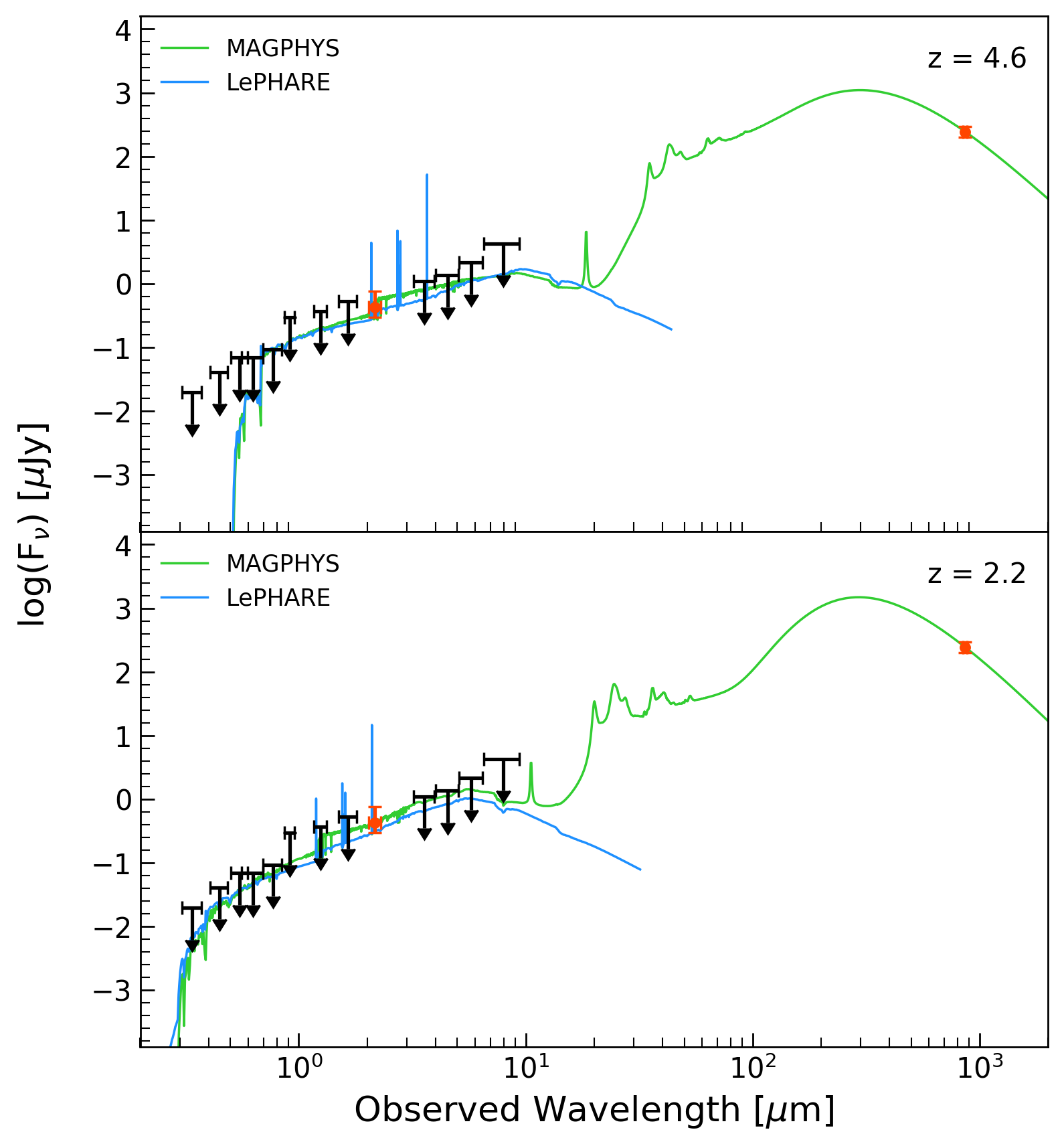}
    \caption{SEDs of Gal-A at $z=4.6$ (top panel) and $z=2.2$ (bottom panel). The green and blue curves are the best-fit models computed with the MAGPHYS and LePHARE codes, respectively. Upper limits on the flux, as reported in Table \ref{tab:phot}, are shown in black. The orange points with the error bars are the detection in the UltraVISTA $K_s$ band and the observed ALMA continuum in Band 7.}
    \label{fig:SED}
\end{center}
\end{figure}

We derive the stellar mass of Gal-A through SED-fitting using LePHARE \citep{Arnouts99, Ilbert06}.

We use a synthetic set of templates of SFGs based on stellar population synthesis models from \cite{Bruzual03}. We explore constant, exponentially declining (with $\tau=0.1,0.3,1,3$ Gyrs) and delayed (with $\tau=0.1,0.5,1,3$ Gyrs) SFHs. To account for metallicity dependence, we use models with solar ($\mathrm{Z_{\odot}}$) and sub-solar (0.2 $\mathrm{Z_{\odot}}$) metallicity. We then account for dust attenuation using the \cite{Calzetti00} attenuation law with a stellar $E_s(B-V)$ ranging from 0 to 0.7 in steps of 0.05. Following \cite{Ilbert09}, we also add the contribution of rest-frame UV and optical emission lines in the different filters. Finally, following \cite{Faisst19}, we perform the fit in flux density space and add systematic errors (depending on the filter) in order to avoid the $\chi^2$ computation to be dominated by small errors.

Fig. \ref{fig:SED} shows the SEDs obtained with LePHARE (blue curves) from the best-fit between the models and the photometry of Gal-A at $z=4.6$ and $z=2.2$. In the first case, the best-fit is given by an exponentially declining model with $\tau=3.0$ Gyrs while, at $z=2.2$, a delayed $\tau$-model with $\tau=0.5$ Gyrs better reproduces the observations.

Since Gal-A is very faint from optical to NIR wavelengths, we decide to perturb the flux in each filter by its relative rms to test the dependence of the fitting on the observed photometry of the galaxy. We thus run a Montecarlo simulation, building 1000 perturbed SEDs that we then refit, in order to obtain a better estimate of the above-mentioned physical parameters from their probability distributions. More in detail, we extract the perturbed flux in each band from a gaussian distribution centred on the measured flux and with standard deviation equal to the measured rms. We list our results in Table \ref{tab:fit}. At $z=4.6$, these results point towards the solution for which Gal-A is a young, dusty SFG. Moreover, as can be seen, the SFR and the FIR luminosity are quite in agreement with the corresponding quantities in Table \ref{tab:param}. Adopting the same procedure for the SED-fitting at $z=2.2$, we find that Gal-A should be a less massive and dustier galaxy with respect to the previous case.

We further compare the results obtained with LePHARE with the MAGPHYS code \citep{daCunha08}, in which we also include the observed ALMA continuum in Band 7. The best models that fit the observations are shown in Fig. \ref{fig:SED} as the green curves. The results from the best-fit, both at $z=2.2$ and $z=4.6$, are very similar to those of LePHARE within the uncertainties. This reassures us about the robustness of our estimates. 

\begin{figure*}
\begin{center}
	\includegraphics[width=12cm]{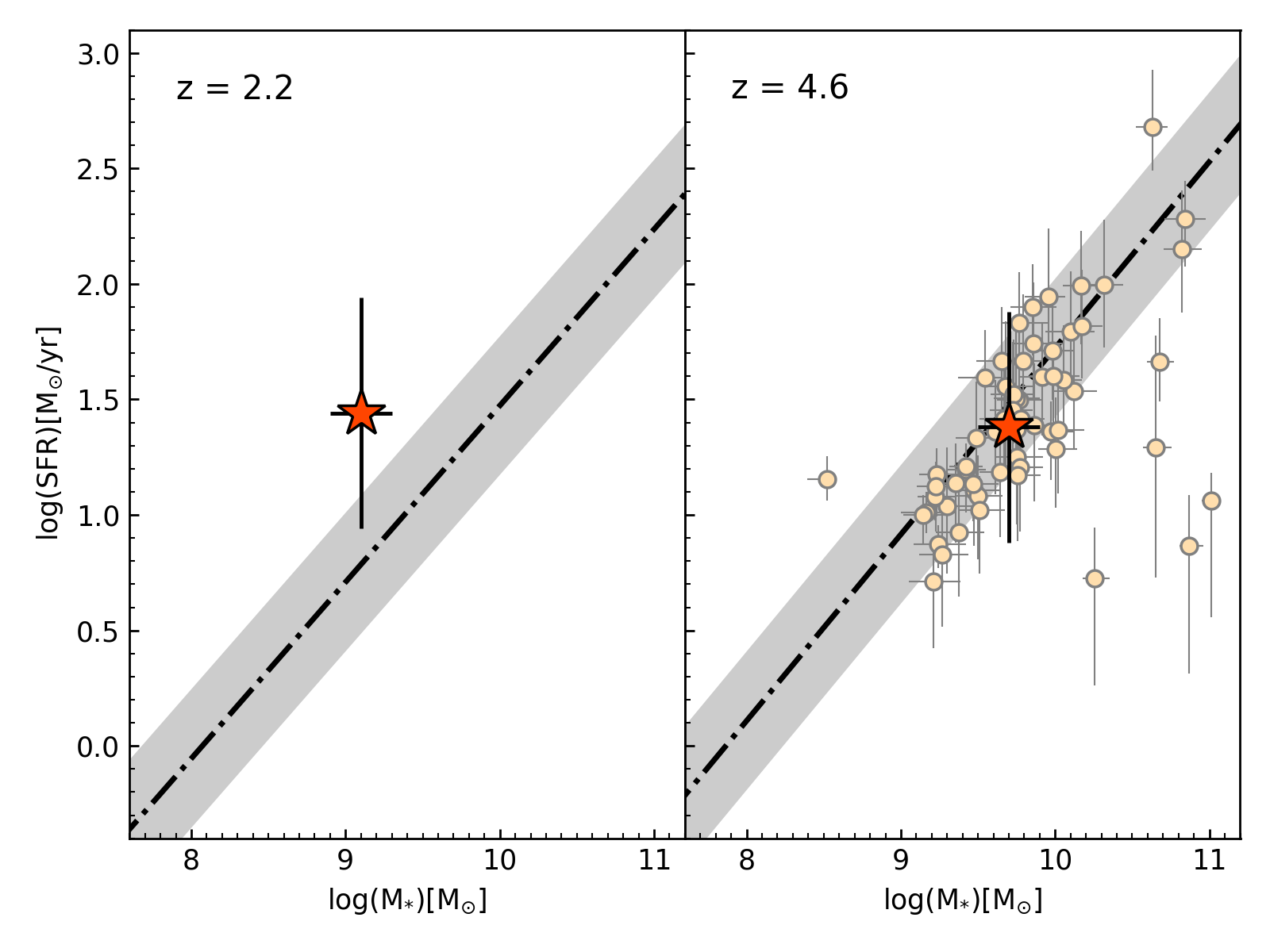}
    \caption{Star-forming MS relations (dot-dashed lines; \citealt{Speagle14}) at redshift 2.2 (left panel) and 4.6 (right panel). The grey bands indicate the scatter from the MS ($\pm0.3$ dex width). The orange stars represent the positions of our source in the diagram, given by the estimated stellar mass from the SEDfitting and the SFR from the FIR luminosity of Gal-A. For the case at $z=4.6$, we also show the positions of the ALPINE galaxies at $4.4\leq z \leq 4.6$ (small circles).}
    \label{fig:mlim}
\end{center}
\end{figure*}

With the stellar mass obtained from the SED-fitting and the SFR measured from the FIR luminosity of the source, we determine the position of Gal-A along the MS of SFGs. In Fig. \ref{fig:mlim} we show the MS relations, assuming a Chabrier IMF, at $z = 2.2$ (left panel) and $z = 4.6$ (right panel) obtained by \cite{Speagle14} combining measurements from previous works in literature. Should the source be at $z=2.2$, it would lie $\sim2\sigma$ above the MS, towards the region populated by starburst galaxies.

Whether the source is at $z=4.6$, instead, it would sit on its corresponding MS. In this case, we also show the location of the ALPINE sample (in the redshift range $4.4\leq z\leq4.6$) in the figure. The ALPINE galaxies have ages in the range $7.8\lesssim \mathrm{log(Age)} \lesssim 9.0$ and $E_s(B-V)$ between 0 and 0.5 \citep{Faisst19}. Gal-A has a similar age to those estimated for the ALPINE targets; moreover, its SFR and $\mathrm{M_{*}}$ are comparable with those of the ALPINE sources and place it along the MS at $z=4.6$. However, the mean $E_s(B-V)$ of the ALPINE galaxies is $\sim0.1$, while Gal-A has $E_s(B-V)\sim0.4$, lying on the tail of the distribution of the color excess, and making it undetected in the optical bands. In this scenario, we should expect an entire population of optically-invisible SFGs, still to be observed, which might significantly contribute to the cosmic SFRD at early times.

\begin{table}
\begin{center}
\begin{tabular}{c c c}
\hline
\hline
Physical & z=2.2 & z=4.6\\
parameters & &\\
\hline
$E_s(B-V)$ & $0.5\pm0.1$ & $0.4\pm0.1$\\
$\mathrm{log(Age/Gyrs)}$ & $7.9\pm0.3$ & $7.8\pm0.1$\\
$\mathrm{log(SFR/M_{\odot}yr^{-1})}$ & $1.3\pm0.4$ & $2.1\pm0.3$\\
$\mathrm{log(L{_{FIR}}/L_{\odot})}$ & $11.0\pm0.1$ & $11.6\pm0.2$\\
$\mathrm{log(M_{*}/M_{\odot})}$ & $9.1\pm0.2$ & $9.7\pm0.2$\\
\hline
\hline
\end{tabular}
\caption{Physical parameters estimated from the SED-fitting at $z=2.2$ and $z=4.6$. Each value represents the mean of the probability distribution obtained perturbing the photometry of Gal-A 1000 times and fitting that photometry with the models. The uncertainties are given by the 16th and 18th percentiles of the distributions.}
\label{tab:fit}
\end{center}
\end{table}

\subsection{Estimate of the dynamical mass}
In this paragraph we attempt an estimate of the galaxy dynamical mass (M$_\mathrm{dyn}$) obtained from the FWHM of the observed emission line. Following \cite{Wang13}, we assume a rotating disk geometry for the gas as a first approximation; in this way, $\mathrm{M_{dyn} = 1.16\times10^5 \hspace{0.5mm} v_{cir}^2 \hspace{0.5mm} D}$, where $\mathrm{v_{cir} = 0.75 \hspace{0.5mm} FWHM_{line}/sin(i)}$ is the circular velocity of the gas disk in km/s (with $i$ the inclination angle between the gas disk and the line of sight), and D is the disk diameter in kpc. Since Gal-A is not resolved, we take the FWHM of the major axis of the 2D Gaussian fitted to the emission line, as the size of our galaxy ($1.06 \pm 0.04$ arcsec, which corresponds to $7.09 \pm 0.27$ kpc at $z \sim 4.6$, and to $8.99 \pm 0.34$ kpc at $z \sim 2.2$). We derive dynamical masses (uncorrected for galaxy inclination) of $\mathrm{M_{dyn} \hspace{0.5mm} sin^2(i) = 4.4 \times 10^{10}}$ $\mathrm{M_{\odot}}$ and $5.6 \times 10^{10}$ $\mathrm{M_{\odot}}$ for $z = 4.6$ and $z=2.2$ respectively, with a 25\% uncertainty obtained from individual errors on the FHWM$_\mathrm{line}$ and on the size of the source. Following \cite{Capak15}, we assume the two values for the inclination angle $\mathrm{sin(i)}=0.45$ and $\mathrm{sin(i)}=1$, ranging from a nearly face-on to an edge-on disk. When $\mathrm{sin(i)=1}$, the previous dynamical masses remain unchanged; however, in the case with $\mathrm{sin(i)=0.45}$, $\mathrm{M_{dyn}}$ increases of a factor 5. This reflects the large uncertainties on the size and geometry of the source, which cannot be well constrained with the current data and our poor resolution. 

Furthermore, this approximation could cease to be valid in case the stellar mass of the source is smaller than the mass threshold above which galaxies are thought to form ordered disks. For instance, \cite{Simons15} found a so-called \lq\lq mass of disk formation\rq\rq{} of $\mathrm{log(M_{*}/M{_\odot})}=9.5$ above which the majority of the galaxies of their sample are rotation-dominated; below this threshold there is instead a large scatter and the galaxies could be either rotation-dominated disks and asymmetric or compact galaxies without any sign of rotation. At $z=2.2$, Gal-A should have $\mathrm{log(M_{*}/M{_\odot})}=9.1$, therefore it is prone to this kind of issue.

For comparison, we also run the 3D-BAROLO algorithm (3D-Based Analysis of Rotating Objects from Line Observations; \citealt{DiTeodoro15}) on the continuum-subtracted data cube to obtain a more accurate estimate of the dynamical mass. This code creates synthetic 3D observations of the galaxy and compares them with the input cube, finding the kinematical and geometrical parameters which best describe the data. It is particularly useful to retrieve information on low-resolution data where the kinematics is biased by the size of the beam, as in this case. We find $\mathrm{log(M_{dyn}/M_{\odot}) = 10.4\pm1.0}$ for $z=4.6$ and $\mathrm{log(M_{dyn}/M_{\odot}) = 10.5\pm1.0}$ for $z=2.2$. These results are quite in agreement with the former, given the large error on $\mathrm{M_{dyn}}$. In particular, at both redshifts, $\mathrm{M_{dyn}/M_{*}} > 1$, likely indicating that the galaxy has recently begun forming stars, resulting in small stellar masses and large gas fractions. However, given the large uncertainties on the dynamical mass, this result is not conclusive.

%%%%%%%%%%%%%%%%%%%%%%%%%%%%%%%%%%%%%%%

\section{Discussion}
In light of the above results, it seems more likely that Gal-A is a dust-obscured galaxy at $z=4.6$. From the [CII]/FIR diagnostic, our source presents similar properties to a large population of SFGs in literature. In addition, the SED-fitting reveals a large dust attenuation as expected for such an obscured galaxy, and places Gal-A along the MS at $z\sim4.6$. Nevertheless, we cannot exclude the possibility that the observed emission line is associated to a dusty, less massive source at $z\sim2.2$, with a $\sim2\sigma$ scatter from the MS and having relatively higher CO luminosities than those typical of local SFGs and high-z sub-millimeter galaxies. In this latter case, the (spectroscopic) redshift of Gal-A would also be comparable with the (photometric) redshifts of Gal-B and Gal-C, maybe suggesting the presence of an on-going merging at that epoch. However, to test this hypothesis, more kinematic information is needed.    

In the most likely scenario in which Gal-A is at $z\sim4.6$, it may be part of the same dark matter halo of DC$\_$665626. In this case, we can assume a stellar mass $-$ halo mass (SMHM) relationship in order to estimate some physical properties of the halo. There are several ways to derive this relation; e.g. \cite{Behroozi10,Behroozi13} used the abundance matching technique to explore the SMHM relation out to $z\sim8$ assuming that the most massive galaxies are monotonically assigned to the most massive halos. Another approach is the Halo Occupation Distribution modeling which assumes that the number of galaxies in a given dark matter halo depends only on the halo mass; \cite{Harikane16} used this method to reproduce the SMHM relation out to $z=7$, obtaining results in agreement with \cite{Behroozi13}. In particular, since Gal-A has a larger stellar mass than DC$\_$665626, we can suppose that the ALPINE target is a satellite galaxy of our serendipitous source embedded in its dark matter halo. In  this case, from the stellar mass of Gal-A (i.e. log$\mathrm{(M_{*}/M_{\odot})} = 9.7 \pm 0.2$), the previously discussed models predict a halo mass between log$\mathrm{(M_{h}/M_{\odot})\sim11.5}$ and log$\mathrm{(M_{h}/M_{\odot})\sim11.7}$. Using the empirical model by \cite{Mashian15}, which links the SFR of the central galaxy to its host halo mass via abundance matching techniques, $\mathrm{M_h}$ also translates into an SFR between $\sim20$ and 40 $\mathrm{M_{\odot} yr^{-1}}$, in agreement, within the uncertainties, with the value estimated from the FIR continuum for Gal-A, i.e. SFR $\sim24$ $\mathrm{M_{\odot} yr^{-1}}$. Exploiting these information and following \cite{Lapi18}, we compute the virial radius of the halo as $\mathrm{R_H \equiv [3 M_H/4\pi \rho_c \Delta_H E_z]^{1/3}}$, where $\mathrm{\rho_c \approx 2.8 \times 10^{11} h^2}$ $\mathrm{M_{\odot}/Mpc^3}$ is the critical density, $\mathrm{\Delta_H \simeq 18 \pi^2 + 82[\Omega_m(1+z)^3/E_z-1] - 39[\Omega_m(1+z)^3/E_z-1]^2}$ is the non-linear density contrast at collapse, and $\mathrm{E_z = \Omega_\Lambda + \Omega_m (1+z)^3}$ is a redshift dependent factor; we obtain $\mathrm{R_H \sim 39-45}$ kpc. Comparing this result to the observed spatial offset between our source and DC$\_$665626 ($\sim 40$ kpc), we may conclude, according to this scenario, that the main ALPINE target could be a low mass satellite in the dark matter halo of Gal-A. 

It is worth noting that we obtain similar results even in the opposite case in which Gal-A is a satellite galaxy of DC$\_$665626. Following the same procedure explained above, and since DC$\_$665626 has $\mathrm{log(M_{*}/M_{\odot}) \sim 9.2}$, we obtain $\mathrm{log(M_{h}/M_{\odot})\sim11.4}$ and $\mathrm{log(SFR/[M_{\odot} yr^{-1}])}\sim1.0$ (which is consistent with the SFR of the ALPINE target obtained through SED-fitting). In turn, this provides $\mathrm{R_H \sim 36}$ kpc, which is again comparable with the observed offset between the two galaxies.

Finally, Gal-A may also be part of the massive proto-cluster of galaxies PCI J1001+0220 located at $z=4.57$ in the COSMOS field \citep{Lemaux18}. In fact, our source lies well inside the 2 Mpc boundary used for spectroscopic membership in that work, with a systemic velocity offset $< 350$ km/s. This strengthens the hypothesis that this source is at $z\sim4.6$. 

%%%%%%%%%%%%%%%%%%%%%%%%%%%%%%%%%%%%%%%

\section{Summary and Conclusions}
In this paper we present the characterisation of Gal-A, a galaxy serendipitously discovered in one of the ALPINE pointings. This source is detected both in line and continuum and does not show any optical counterpart, from UV to FIR, except for the $K_s$ band from UltraVISTA (DR4). This leads to high uncertainties on the real nature of the observed emission line, i.e. [CII] at $z_\mathrm{gal} = 4.577$, CO(9-8) or CO(10-9) at $z_\mathrm{gal} = 2.043$ and 2.381, respectively.  

Although we cannot definitively exclude that Gal-A is a dust-obscured galaxy at $z\sim2.2$, the analysis undertaken in this work suggests that this source is more likely a $z\sim4.6$ MS SFG missed by UV/optical surveys because of its high level of dust-obscuration. Moreover, at this epoch, several dusty galaxies without optical/NIR detections have been yet confirmed, mostly as extreme starbursts (e.g. \citealt{Riechers13,Riechers17,Alcalde19}); Gal-A could be part of this elusive population of sources, with a smaller luminosity and/or mass. In this last case, we compute an SFR $\sim 24$ $\mathrm{M_{\odot}/yr}$ from the FIR luminosity, $\mathrm{log(M_{*}/M_{\odot})} \sim 9.7$ and an age of $\sim70$ Myrs from Montecarlo simulations on the SED-fitting procedure.

Whether the emission comes from CO or [CII], both the cases presented above are undoubtedly interesting. If it was at $z\sim2.2$, our galaxy would increase the sample of high-J CO emitters at high redshift, leading to a more in-depth study of the excitation conditions of the molecular gas in these sources; only a handful of these kind of objects has been detected so far, and most of them seem to be associated with active galactic nuclei (AGNs) activities \citep{Weiss07,Riechers11,Riechers13}. Should the serendipitous emission be [CII] instead, we would identify a SFG invisible to optical-NIR observations. [CII] emission traces recent star formation in the galaxy, and the ALPINE survey will allow us to quantify how many similar objects to the one analysed in this work we will be able to discover. In fact, among the serendipitous sources found in ALPINE, there is a high fraction of objects without UV/optical counterparts (Loiacono et al. in prep.). Thanks to ALPINE, we are now able to estimate the overall contribution of these dust-obscured galaxies to the SFRD in the early Universe.   

Eventually, we plan to spectroscopic follow-up this source in order to firmly establish the nature of its emission line. For instance, ALMA observations in Band 6 could reveal [NII] emission at 205 $\mu$m rest-frame if the galaxy is at $z\sim4.6$; in this case the ratio [CII]/[NII] would also provide the fraction of [CII] emission arising from the ionised gas, i.e. from star-forming regions \citep{Oberst06,Oberst11,Zhao16}. X-shooter at the Very Large Telescope (VLT) could also be useful to unveil the redshift of this source by observing [OII] emission at $z=4.6$, or even H$\alpha$ emission redshifted in the NIR region of the spectrum at $z\sim2.2$. However, these observations could be hampered by the large $E_s(B-V)$ found for this source which makes it invisible in optical filters. Finally, the Near-Infrared Spectrograph (NIRSpec) on the James Webb Space Telescope (JWST) will be a powerful facility for the follow-up of this kind of sources as well.

\section*{Acknowledgements}
This paper is based on data obtained with the ALMA observatory, under the Large Program 2017.1.00428.L. ALMA is a partnership of ESO (representing its member states), NSF (USA) and NINS (Japan), together with NRC (Canada), MOST and ASIAA (Taiwan), and KASI (Republic of Korea), in cooperation with the Republic of Chile. The Joint ALMA Observatory is operated by ESO, AUI/NRAO and NAOJ. Based on data products from observations made with ESO Telescopes at the La Silla Paranal Observatory under ESO programme ID 179.A-2005 and on data products produced by CALET and the Cambridge Astronomy Survey Unit on behalf of the UltraVISTA consortium. This work has made use of the Rainbow Cosmological Surveys Database, which is operated by the Centro de Astrobiología (CAB/INTA), partnered with the University of California Observatories at Santa Cruz (UCO/Lick,UCSC). We thank D. Liu and collaborators for providing us individual values of CO and FIR luminosities estimated in their work. S.B., A.C., C.G., F.L., F.P., G.R., and M.T. acknowledge the support from grant PRIN MIUR 2017. L.V. acknowledges funding from the European Union’s Horizon 2020 research and innovation program under the Marie Sk lodowska-Curie Grant agreement No. 746119. D.R. acknowledges support from the National Science Foundation under grant numbers AST-1614213 and AST-1910107 and from the Alexander von Humboldt Foundation through a Humboldt Research Fellowship for Experienced Researchers. G.C.J. acknowledges ERC Advanced Grant 695671 ``QUENCH'' and support by the Science and Technology Facilities Council (STFC). S.F. is supported by the Cosmic Dawn Center of Excellence founded by the Danish National Research Foundation under the grant No. 140.

%%%%%%%%%%%%%%%%%%%%%%%%%%%%%%%%%%%%%%%%%%%%%%%%%%

%%%%%%%%%%%%%%%%%%%% REFERENCES %%%%%%%%%%%%%%%%%%

% The best way to enter references is to use BibTeX:

%\bibliographystyle{mnras}
%\bibliography{example} % if your bibtex file is called example.bib

% Alternatively you could enter them by hand, like this:
% This method is tedious and prone to error if you have lots of references

%%%%%%%%%%%%%%%%%%%%%%%%%%%%%%%%%%%%%%%%%%%%%%%%%%

%%%%%%%%%%%%%%%%% APPENDICES %%%%%%%%%%%%%%%%%%%%%

%%%%%%%%%%%%%%%%%%%%%%%%%%%%%%%%%%%%%%%%%%%%%%%%%%

% Don't change these lines
\bsp	% typesetting comment
\label{lastpage}

\section*{Affiliations}
% List of institutions
{\it
$^{1}$Dipartimento di Fisica e Astronomia, Universit\`a di Padova, Vicolo dell'Osservatorio 3, I-35122, Padova, Italy\\
$^{2}$INAF--Osservatorio Astronomico di Padova, Vicolo dell'Osservatorio 5, I-35122, Padova, Italy\\
$^{3}$Department of Physics, University of California, Davis, One Shields Ave., Davis, CA 95616, USA\\
$^{4}$Aix Marseille Univ, CNRS, CNES, LAM, Marseille, France\\
$^{5}$IPAC, California Institute of Technology, 1200 East California Boulevard, Pasadena, CA 91125, USA\\
$^{6}$The Cosmic Dawn Center, University of Copenhagen, Vibenshuset, Lyngbyvej 2, DK-2100 Copenhagen, Denmark\\
$^{7}$Niels Bohr Institute, University of Copenhagen, Lyngbyvej 2, DK-2100 Copenhagen, Denmark\\
$^{8}$Observatoire de Genève, Universit\'e de Genève 51 Ch. des Maillettes, 1290 Versoix, Switzerland\\
$^{9}$Institut de Recherche en Astrophysique et Plan\'etologie - IRAP, CNRS, Universit\'e de Toulouse, UPS-OMP, 14, avenue E. Belin,\\
$^{10}$Kavli Institute for the Physics and Mathematics of the Universe, The University of Tokyo Kashiwa, Chiba 277-8583, Japan\\
$^{11}$Department of Astronomy, School of Science, The University of Tokyo, 7-3-1 Hongo, Bunkyo, Tokyo 113-0033, Japan\\
$^{12}$The Caltech Optical Observatories, California Institute of Technology, Pasadena, CA 91125, USA\\
$^{13}$Osservatorio di Astrofisica e Scienza dello Spazio - Istituto Nazionale di Astrofisica, via Gobetti 93/3, I-40129, Bologna, Italy\\
$^{14}$Centro de Astronom\'ia (CITEVA), Universidad de Antofagasta, Avenida Angamos 601, Antofagasta, Chile\\
$^{15}$University of Bologna, Department of Physics and Astronomy (DIFA), Via Gobetti 93/2, I-40129, Bologna, Italy\\
$^{16}$INAF - Osservatorio Astrofisico di Arcetri,
Largo E. Fermi 5, I-50125, Firenze, Italy\\
%$^{17}$Research Institute for Science and Engineering, Waseda University,3-4-1 Okubo, Shinjuku, Tokyo 169-8555, Japan\\
$^{17}$Niels Bohr Institute, University of Copenhagen, Lyngbyvej 2, DK-2100 Copenhagen, Denmark\\
%$^{18}$National Astronomical Observatory of Japan, 2-21-1, Osawa, Mitaka, Tokyo, Japan\\
$^{18}$Space Telescope Science Institute, 3700 San Martin Drive, Baltimore, MD 21218, USA\\
$^{19}$Instituto de Física y Astronomía, Universidad de Valparaíso, Avda. Gran Bretaña 1111, Valparaíso, Chile\\
$^{20}$Cavendish Laboratory, University of Cambridge, 19 J. J. Thomson Ave., Cambridge CB3 0HE, UK\\
$^{21}$Kavli Institute for Cosmology, University of Cambridge, Madingley Road, Cambridge CB3 0HA, UK\\
$^{22}$Department of Astronomy, Cornell University, Space Sciences Building, Ithaca, NY 14853, USA\\
$^{23}$Max-Planck-Institut f\"ur Astronomie, K\"onigstuhl 17, D-69117, Heidelberg, Germany\\
$^{24}$Leiden Observatory, Leiden University, PO Box 9500, 2300 RA Leiden, The Netherlands\\
}
\end{document}